\documentclass[aps,twocolumn,showpacs,hyperref]{revtex4-1}
\usepackage{graphicx}
\usepackage{subfigure}
\usepackage{bm}
\usepackage{float}

\newcommand{\taubf}{$\tau_{bf}$}
\newcommand{\taual}{$\tau_{al}$}
\newcommand{\rb}{$C_{bf}$}
\newcommand{\rbcr}{$C^{cr}_{bf}$}

\begin{document}
\title{Achieving Control of Lesion Growth in CNS with Minimal Damage}
\author{Mathankumar Raja}
\email{mathankraja@gmail.com}
\author{Krishna Mohan, T. R.}
\email{kmohan@cmmacs.ernet.in}
\homepage{http://www.cmmacs.ernet.in/~kmohan}
\affiliation{CSIR Centre for Mathematical Modelling and Computer Simulation (C-MMACS)\\Bangalore 560017, India}
\date{\today}

\begin{abstract}
Lesions in central nervous system (CNS) and their growth leads to debilitating diseases like Multiple Sclerosis (MS), Alzheimer's etc. We developed a model earlier \cite{orgin1,orgin2} which shows how the lesion growth can be arrested through a beneficial auto-immune mechanism. The success of the approach depends on a set of control parameters and their phase space was shown to have a smooth manifold separating the uncontrolled lesion growth region from the controlled. Here we show that an optimal set of parameter values exist which minimizes system damage while achieving control of lesion growth.
\end{abstract}

\maketitle 

\section{Introduction}
Human brain normally contains more than eight billion neurons and, with their respective axonal connections, create a huge complex network. Multiple sclerosis (MS) is a disease of active demyelination of neurons with destruction of brain functionality \cite{mye}. MS directly affects the connectivity of the affected region by disrupting signal transmission but the pathogenetic mechanism of MS is still unclear. Various clinical studies have been attempting to explain the pathophysiology of the disease. Some studies proposed MS as a neurodegenerative disorder where our own immune system, fooled perhaps by molecular mimicry, causes destruction of own nervous system \cite{msimmune}.

Clinical pathology studies have demonstrated heterogeneity of MS in the immunopathological profiles of the lesions in different cases. Lesions are categorized as autoimmune encephalomyelitis (patterns I and II) or oligodendrocyte dystrophy (patterns III and IV) using a broad spectrum of immunological and neurological markers \cite{hetropattern}. In MS with pattern III lesions, demyelination is induced by a functional loss of oligodendrocytes, possibly as a result of infection with an unknown virus. European Union FP6 NeuroproMiSe project study provides the evidence for an association between Epstein virus infection and MS, but unequivocal proof of the same is still lacking \cite{virusbrain}.

In our previous work, we invoked apoptosis initiated in peripheral regions of infection by an intercellular distress signaling mechanism to block  pathogen advance. This model resembles the action of firemen who control wild fire by burning peripheral vegetation. Our model demonstrated that spread of the infection can be contained by such a systematic immune response initiated in peripheral regions of infection \cite{orgin1,orgin2}.

An alternative model supports the possibility of tissue pre-conditioning in the same peripheral layers as a method to control the radial spread of lesions \cite{precond}. In such a mechanism, stress signals emanating from the pathogenically affected cells, just as in the apoptosis model above, induce expression of heat shock proteins and hypoxia-inducible factors in the cells of the rim region of lesions. Due to the neuroprotective effects of the expressed proteins on the tissues, cells are strengthened to resist further infection. It is possible that both of the above approaches are employed in the real systems \cite{precond}.

Lesion growth affects structural connectivity of the neural network and results in losses in functionalities. We adopt graph-theoretic techniques in this paper to quantify the network robustness and loss of functionalities in the complex brain networks. We also establish here that, while a range of values of control parameters enables control of infection, an optimal set exists, which will minimize damage to the system.

\section{Model}
Extraction of complete axonal projection map of all neurons of mammalian brain is currently out of reach \cite{connect} and, so, axonal connectivity of the CNS is here represented by a fixed radius random graph. This consists of nodes generated randomly in a unit square with connections established between each node and all its neighbors within a fixed radius R. Initially, health status of all edges (axons) are assigned an arbitrarily assigned maximum value (set to 1 here) and these weights become zero when they are killed by infection or apoptosis. Pathogenic process is driven by probabilistic events with a pre-assigned probability value of damaging edges in each visit to an edge.

Damages to its edges at a particular node triggers an alarm signal when the health status of the node falls below a threshold, \taual. This stress signal is propagated to all connected nodes from where it is propagated further. Apoptosis process is initiated at a node when accumulated alarm signals at the node reaches a threshold, \taubf.

Apoptosis affects a circular region with a radius proportional to a parameter, \rb, and a higher value of \rb\ implies apoptosis of a larger region. In apoptosis, a circular region around the activated node  gets reset to zero. No additional signals are generated at these nodes to the alarm signals generated in the pathological process.

\begin{figure*}
\begin{center}
\includegraphics[width=\linewidth,clip]{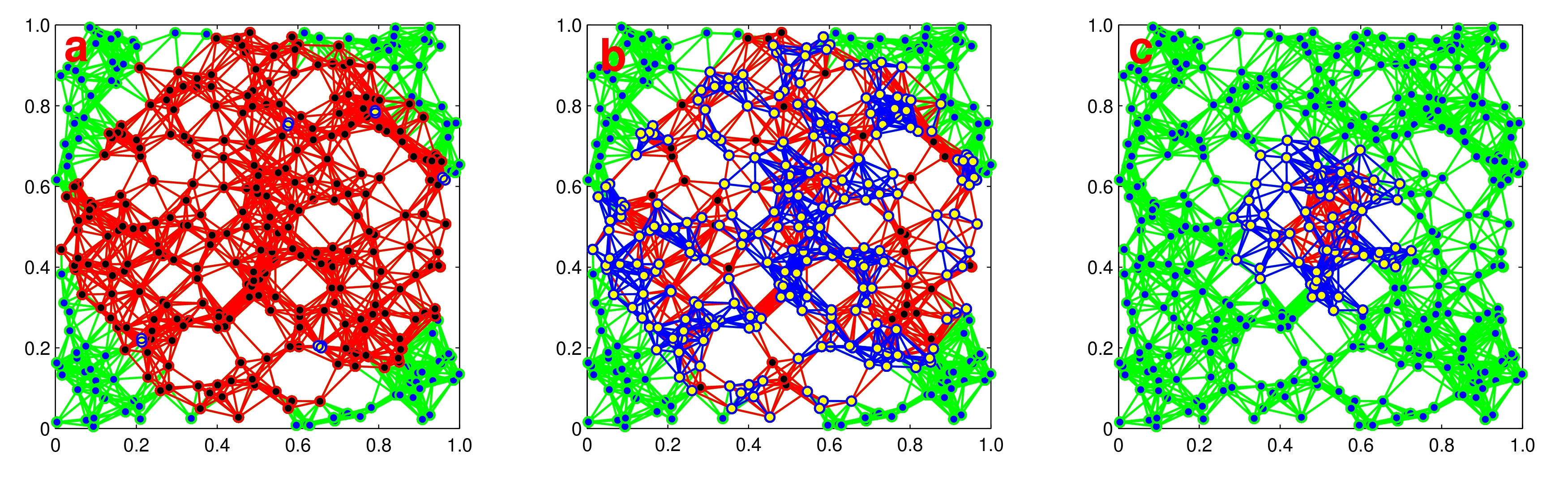}
\caption{Snapshots of the network for three different cases of lesion dynamics. (a) Uncontrolled infection (red) killed most of the healthy edges (green) because insufficient apoptosis process (blue) failed to check it (\taual=0.5,  \rb=0.1 and  \taubf=0.5). (b) War between apoptosis process and infection is on, but the intensity of apoptosis process is not enough to check the infection. On the contrary, it only succeeds in adding to total damage (\taual=0.5, \rb=0.8, and \taubf=0.5). (c) Success of infection control attained with minimal damage by the efficient action of apoptosis process (\taual=0.5, \rb=1.5 and \taubf=0.5).}
\label{net}
\end{center}
\end{figure*}

Fig.~\ref{net} gives snapshots of three different possble scenarios. In Fig.~\ref{net}a, when the control level is low, infection is seen to consume the entire system. In Fig.~\ref{net}b, level of control is still insufficient and this only aids to spread the damage further because apoptosis also adds to the killed edges. In Fig.~\ref{net}c, we finally have the right level of control and infection is well arrested.

In strengthening process, unlike in apoptosis, weightage of edges in the circular region controlled by \rb\ is increased to resist against infection. We assume here that, once strengthened, those edges become immune to further infection. A detailed view of our model can be had from our previous publications \cite{orgin1,orgin2}.

\subsection*{Network Measures}
The status of the system is encoded into an adjacency matrix and quantitative study of loss of functionalities due to the infection and associated system protective reactions can be explored through parameters that show network quality such as these listed below.

{\bf Flow Coefficient: } It is a measure of ``local centrality'' to estimate the capacity of a node to conduct information flow between its neighboring nodes. It is calculated as the number of actual paths of length two divided by the number of all possible paths of length two that traverse a central node \cite{flowhoney2007}.

{\bf Global Efficiency:} Path length is defined the minimum number of edges that must be traversed to go from one node to another. Global efficiency is related to average inverse shortest path length and is inversely related to characteristic path length. It is numerically easier to compute than characteristic path length and is useful to estimate topological distance between elements of disconnected graphs \cite{eff2001}.

{\bf Edge Density: }It is the proportion of connections that exists relative to the number of potential connections of a network. An edge density of 1, corresponding to a percentage of 100\%, would mean that all possible edges exist  \cite{density}.

{\bf Total Wiring Length:} It is the total number of walks through all possible paths that never visit the same path again. One recent study evaluated importance of the optimal wiring length in brain network projections which makes this measure important for evaluating connectivity \cite{wirelength}.

 Snapshot of the dynamical events are taken at each time step and stored in the form of adjacency matrices which are then analyzed to estimate values of the above parameters. Identifying the quality of the complex networks needs a multi-level approach with different indices such as those listed above.

\section {Results}
As in our previous study, health status of the edges are either alive (=1) or dead (=0) and probability of pathologic damage was set to be  $p_d = 0.33$. Number of nodes was set to 400 and pathological damage was initiated at the center of a unit square. We aimed to determine the total damage to the network for different value of the three control parameters (\taual, \taubf\ and \rb) under both strengthening and apoptosis control methods.
 
In our previous studies, we indicated that the pathological process is always controlled if \rb\ value lies above a critical value, \rbcr(\taual, \taubf). We had also postulated that damage to the system will increase with subsequent increase in \rb\ due to the increased aggressiveness of apoptosis (see our Fig.~3 in \cite{orgin2}). We illustrate this aspect below and focus here on the infection-controlled cases for a series of different parameter values in (\taual,\taubf,\rb) to identify the optimal values in the parametric space which aids us to minimize the damage to the system. 

\begin{center}
\begin{figure*}
\includegraphics[width=\linewidth,clip]{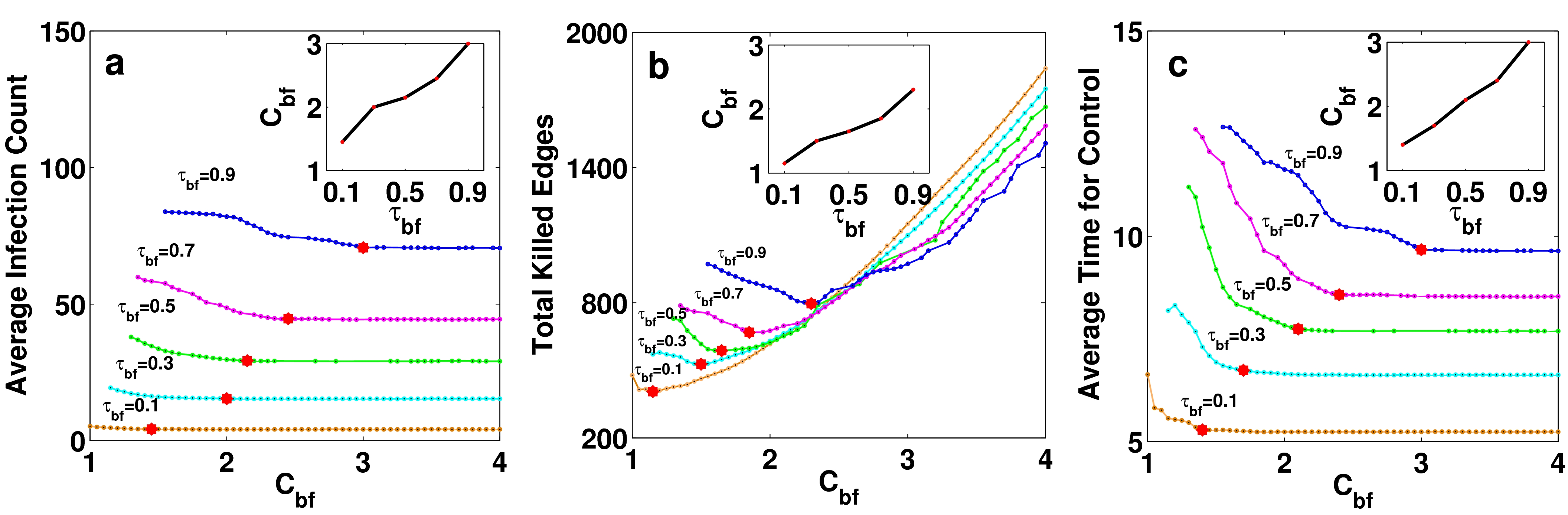}
\caption{(a) Variation in number of edges damaged by infection  with increase in \rb\ values for different \taubf\ values (indicated in figure) and a fixed \taual=0.5.  (b) Variation in total damage (due to edges killed by infection as well as apoptosis) of the system as \rb\ is increased for fixed \taual= 0.5 and different \taubf\ values (indicated in figure). (c) Variation in  time taken to control infection as \rb\ is increased for fixed \taual= 0.5 and different \taubf\ values (indicated in figure). The insets to the figures plot the values of \rb\   at the minimum of the graphs versus the respective \taubf\  values.}
\label{all}
\end{figure*}
\end{center}

Fig.~\ref{all}(a) shows that, if we monitor the total number of edges damaged by infection after it has been controlled, we find it to progressively reduce as \rb\ is increased from 1 to 3  (lesser values when \taubf\ is smaller). This continues till a level of  minimum damage is attained, after which no further reduction is possible however much we increase \rb. This is what we had anticipated in an earlier  work. On the contrary, the total damage to the system increases for further increase in \rb. This is seen in Fig.~\ref{all}(b). This growth in total damage is due to the unnecessary damage caused by the apoptosis process. Increase in \rb\ leads to larger areas being subjected to apoptosis in each apoptotic event and this is clearly unnecessary in this case. Fig.~\ref{all}(c) strengthens our argument. It is seen here that the time taken to control infection attains its minimum for the same values of \rb\ where minimum damage due to infection occurs as well as the minimum in total damage occurs. 

As we can see in the insets to Fig.~\ref{all}, \rb\ value increases approximately linearly  with \taubf. This happens because when \taubf\ increases, the reduced sensitivity of the alarm signal causes apoptosis process to be initiated with a delay and, it then takes longer to control the infection. A similar situation occurs when \taubf\ is held fixed and \taual\ is allowed to vary. However, in this case, smaller values of \taual\ require larger values of \rb\ to achieve control of infection (see~\cite{orgin2}).

In the tissue preconditioning process, a very similar scenario results as in the above. However in this case, we do not see an increase in total damage for values of  ${C_{bf} > C_{bf}^{cr}}$ because the total damage is all due to infection alone and, when the infection process is controlled, there is no more damage because of the absence of the apoptosis process. In biological systems, it is possible that a control system to bring to a halt, when needed, the apoptosis process, as also the tissue preconditioning process, exists which will take care of unnecessary damages or wastage of energy. In any case, our basic message is that an optimal set of parameter values exist which minimizes damage and wasteful energy leakage.

\begin{center} 
\begin{figure}
\includegraphics[width=\linewidth,clip]{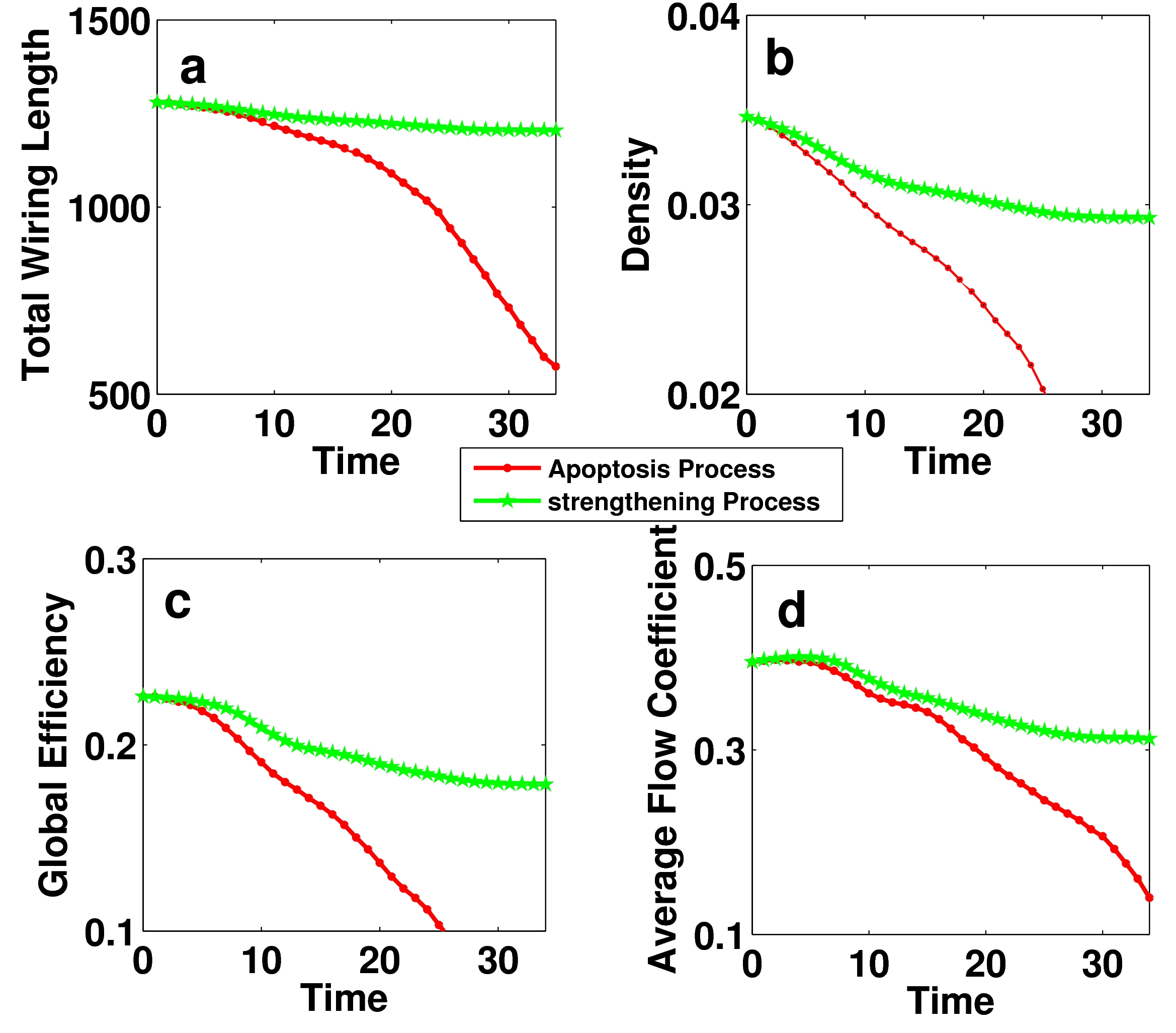}
\caption{Variation in different indices denoting quality of network as the dynamics between infection and its control through apoptosis/strengthening process progress towards the final state of controlled infection. The graph is typical across all values in the ranges of the control parameters and, is depicted here for \taual=0.6, \rb=1.6 and \taubf=0.6.}
\label{para}
\end{figure}
\end{center}
In Fig.~\ref{para}, the evolution of some parameters indicating network quality are shown, through the processes of infection and the ensuing reaction from the immune system. The values of the parameters are shown for both the apoptosis and tissue preconditioning approaches. It is seen that the quality of the network is preserved better in the tissue strengthening process, as is to be expected,  than the apoptosis process.

\section {Conclusion}
In an earlier study~\cite{orgin1, orgin2}, modeling of MS infection process, or equivalently just the lesion growth in CNS, and its possible control by the immune system,   was carried out employing a fixed radius network and signaling mechanisms which initiate apoptosis in regions surrounding the infected regions. Akin to firemen who burn peripheral vegetation to control spread of wild fires, the apoptosis process was shown to be capable of terminating the lesion growth.  Our approach  showed that autoimmunity can work beneficially and need not be considered to be a defect in the system. In this we were following earlier work by Matzinger et al. and others~\cite{danger}.

In this study, we have showed that an optimal set of values exist for the control parameters (\taual, \taubf, \rb) which minimizes damage to the system while at once controlling the infection.  We have not tried to model the process of stopping the apoptosis process after successful control of infection has been achieved. Neither have we attempted to model how the system may arrive at the optimal values of the control parameters. These aspects can be studied after the biological elements which can implement the above model of autoimmunity have been identified. Currently, implementation of our beneficial autoimmune model to the real human brain connectivity network is being undertaken to overcome limitations of the fixed radius model as also to bring the model into a realistic environment where identification of biological elements that effect the autoimmune response can be attempted.

\bibliographystyle{plain}
\bibliography{refbib}
\end{document}